\documentclass[12pt]{article}

\begin{document}

\title{ Optical vortices in dispersive nonlinear Kerr type media}

\author{Lubomir M. Kovachev\\
Institute of Electronics, Bulgarian Academy of Sciences,\\
Tzarigradcko shossee 72,1784 Sofia, Bulgaria}
\date{14 April 2003}
\maketitle
\begin{abstract}

The applied method of slowly varying amplitudes of the electrical
and magnet vector fields give us the possibility to reduce the
nonlinear vector integro-differential wave equation to the
amplitude vector nonlinear differential equations. It can be
estimated different orders of dispersion of the linear and
nonlinear susceptibility using this approximation. The critical
values of parameters to observe different linear and nonlinear
effects are determinate. The obtained amplitude equation is a
vector version of 3D+1 Nonlinear Schredinger Equation (VNSE)
describing the evolution of slowly varying amplitudes of
electrical and magnet fields in dispersive nonlinear Kerr type
media. We show that VNSE admit exact vortex solutions with
classical orbital momentum $\ell=1$ and finite energy. Dispersion
region and medium parameters necessary for experimental
observation of these vortices, are determinate.

\end{abstract}

\section{Introduction}

In present day, there are no difficulties for experimentalist in
laser physics or nonlinear optics to obtain picosecond or
femtosecond optical pulses with equal duration in $x$, $y$ and $z$
directions. The problems with so generated light bullets arise in
the process of their propagation in a nonlinear media with
dispersion.  In the transparency region of a dispersive Kerr type
media, as it was established in \cite{KEL, KARP, MN}, the scalar
paraxial approximation (no dispersion in the direction of
propagation), is in very good accordance with the experimental
results. The paraxial approximation, used in the derivation of the
scalar 2D+1 nonlinear Schredinger equation (NSE), do not include a
small (in first or second order of magnitude) second derivative of
the amplitude function in the direction of propagation. The
including of this term does not change the main result
dramatically: in the case of a linear propagation (pulses with
small intensity), as generated optical bullets at short distance
are transformed in optical disks, with large transverse and small
lengthwise dimension. Some of the experimental possibilities, this
small term to become important, were discussed in \cite {JMN}. In
Ref.\cite {AK} it was shown, that this second derivative in the
direction of propagation term is in the same order and with the
same sign as the others only in some special cases: optical pulses
near the Langmuir frequency or near some of the electronic
resonances. In these regions the sign of the dispersion is
negative and the scalar 2D+1 NSE becomes 3D+1 NSE ones
\cite{SIL,EE,MWBL}. Propagation of optical bullets under the
dynamics of 2D+1 and 3D+1 NSE are investigated also in relation to
different kind of nonlinearity \cite{AHM, EDE}.
      A generation of a new kind of 2D and 3D optical pulses, so called optical
vortices, has recently become a topic of considerable interest.
Generally, the optical vortices are such type of optical pulses,
which admit angular dependence of electrical field or helical
phase distribution. The electrical field or intensity is zero also
in the center of the vortices. The original scalar theory of
optical vortices was based on the well known 2D+1 NSE \cite{SW,
CRIS, STA}. In a self focusing regime of propagation can be
generated optical rings, but they are modulationally unstable
\cite{MAM, TCRIS}. One alternative way of stabilizing of optical
vortices in 2D and 3D case, using saturable \cite{TOR, SKR} or
cubic- quintic \cite{DESM, MAL} nonlinearity, was discussed also.
From the other hand, the experiments with optical vortices show,
that the polarization, and the vector character of the electric
field, play  an important role in the dynamics and the
stabilization of the vortices \cite{LAW}. To investigate these
cases we are going to a vector version of 3D+1 NSE. In Section 2
we derive this vector version of 3D+1 NSE (VNSE), using also the
dispersion in the direction of propagation and envelope
approximation in the standard way as it was constructed in
\cite{KARP,MN}. This vector generalization allows us to find exact
vortex solutions with spin l=1 and finite energy. The question for
shape and finiteness energy of these vortices is crucial, and will
discuss largely in Section 5.

\section{Maxwell's equations for a source-free, dispersive, nonlinear
 Kerr type medium}

      The Maxwell's equations for a source-free dispersive media with Kerr
 nonlinearity are:

\begin{eqnarray}
\label{eq1}
\nabla \times \vec E =
- \frac{1}{c}\frac{\partial \vec B}{\partial t},
\end{eqnarray}

\begin{eqnarray}
\label{eq2}
\nabla \times \vec H =
  \frac{1}{c}\frac{\partial \vec D}{\partial t},
\end{eqnarray}

\begin{equation}
\label{eq3}
\nabla \cdot \vec D = 0,
\end{equation}

\begin{equation}
\label{eq4}
\nabla \cdot \vec B = \nabla \cdot \vec H = 0,
\end{equation}

\begin{equation}
\label{eq5}
\vec B = \vec H,\
\vec D = \vec P_{lin}  + 4\pi \vec P_{nlin},
\end{equation}
where $\vec E$ and $\vec H$ are the electric and magnetic
intensity fields, $\vec D$ and $\vec B$ are the electric and
magnetic induction fields, $\vec P_{lin}$, $\vec P_{nlin}$ are the
linear and nonlinear polarizations of the medium respectively. The
linear electric polarization of an isotropic, disperse medium can
be represented as:

\begin{eqnarray}
\label{eq6} \vec P_{lin} = \int\limits_{-\infty}^{t} {\left(\delta
(t-\tau) +4\pi\chi^{\left(1\right)} \left(t-\tau\right)\right)\vec
E\left(\tau, x, y, z\right)}d\tau =\nonumber\\
\int\limits_{-\infty}^{t} {\varepsilon _0\left(t-\tau\right)\vec
E\left(\tau, x, y, z\right)}d\tau,
\end{eqnarray}
where $\chi^{(1)}$ and $\varepsilon _0$ are the linear electric
susceptibility and the dielectric constant. In the following, we
will study such polarization (linearly or circularly polarized
light), where the nonlinear polarization in the case of one
carrying frequency can be expressed as:

\begin{eqnarray}
\label{pnl} \vec P_{nl}^{(3)} =
\int\limits_{-\infty}^{t}\int\limits_{-\infty}^{t}\int\limits_{-\infty}^{t}
\chi^{\left(3\right)}
\left(t-\tau_1,t-\tau_2,t-\tau_3\right)\nonumber\\
\times\left(\vec{E}(\tau_1,x,y,z)\cdot\vec
E^{*}(\tau_2,x,y,z)\right)\vec E(\tau_3,x,y,z)
d\tau_1d\tau_2d\tau_3.
\end{eqnarray}
It is important to point here the remark of Akhmanov at all in
\cite{AVC}, that  the method slowly varying amplitudes of
electrical and magnet fields applied to such nonstationary linear
and nonlinear representation is valid as well when the optical
pulse duration of the pulses $t_0$ is greater than the
characteristic response time of the media $\tau_0$
($t_0>>\tau_0$), as when the time duration of the pulses are equal
or less than the time response of the media ($t_0\leq\tau_0$).
This possibility is discussed in the process of deriving of the
amplitude equations.

Taking the curl of equation (\ref{eq1}) and using (\ref{eq2}) and
(\ref{eq5}), we then obtain:

\begin{eqnarray}
\label {eq8}
\Delta\vec E-\nabla \left(\nabla\cdot\vec E\right) -
\frac{1}{c^2}\frac{\partial^2\vec D}{\partial t^2}=0,
\end{eqnarray}
where $\Delta \equiv \nabla ^2$ is the Laplasian operator.
Equation (\ref{eq8}) is derived without the use of the third
Maxwell's equation. Using equation (\ref{eq3}) and the expression
for the linear and nonlinear polarizations (\ref{eq6}) and
(\ref{pnl}), it can be estimate the second term in the equation
(\ref{eq8}) for arbitrary localized vector function of the
electrical field. As is shown in \cite{AK}, for such type of
function $\nabla \cdot\vec {E}\cong 0$ and equation (8) is
written:

\begin{eqnarray}
\label{eq9}
\Delta \vec E - \frac{1}{c^2}\frac{\partial^2\vec D}{\partial t^2}= 0.
\end{eqnarray}
We define a complex presentation of the real electrical field by
the relation:

\begin{eqnarray}
\label{a1}
 \vec E\left(x,y,z,t\right)=
\vec {A}'\left(x,y,z,t\right)
\frac{1}{2i}\left(\exp{\left(i(k_0z-\omega_0t)\right)}-c.c\right),
\end{eqnarray}
where $\vec A' $ and $\omega_0$ and $k_0 $ are the real vector
amplitude, the optical frequency and the wave number of the
optical field respectively. The real vector amplitude $\vec A' $
is represented also by a complex vector amplitude field:

\begin{eqnarray} \label{qq1}
 \vec A'\left(x,y,z,t\right)=
\frac{1}{2i}\left(\vec A\left(x,y,z,t\right)-\vec
A^*\left(x,y,z,t\right)\right),
\end{eqnarray}
where $\vec A$ is the complex amplitude of the electrical field.
In this way our real electrical field is presented with four
complex fields of kind:

\begin{eqnarray}
\label{xx1} \vec {E}\left(x,y,z,t\right)= -\frac {1}{4}\left(\vec
{A}\exp{(i(k_0z-\omega_0t))}+\vec
{A}^*\exp{(-i(k_0z-\omega_0t))}\right)\nonumber\\
+\frac {1}{4}\left(\vec{A}\exp{(-i(k_0z-\omega_0t))}+\vec
{A}^*\exp{(i(k_0z-\omega_0t))}\right).
\end{eqnarray}
This special kind of presentation of electrical field is connected
with shape and finiteness energy of vortex solutions, and it is
discussed in Section 5. Here are used also the Fourier
representation of the complex amplitude function $\vec A$ and
their time derivatives to second order:

\begin{eqnarray}
\label {a2} \vec A\left(x,y,z,t\right)=
\int\limits_{-\infty}^{+\infty} \vec
A\left(x,y,z,\omega-\omega_0\right)
\exp{\left(-i\left(\omega_0-\omega\right)t\right)}d\omega,
\end{eqnarray}

\begin{eqnarray}
\label {a3} \frac{\partial\vec A\left(x,y,z,t\right)}{\partial t}
= -i\int\limits_{-\infty}^{+\infty}
    \left(\omega-\omega_0\right)
    \vec A\left(x, y, z,\omega-\omega_0\right)\nonumber\\
\times \exp{\left(-i\left(\omega-\omega_0\right)t\right)}d\omega,
\end{eqnarray}

\begin{eqnarray}
\label {a4}
 \frac{\partial^2\vec A\left(x,y,z,t\right)}{\partial t^2}
= -\int\limits_{-\infty}^{+\infty}
   \left(\omega-\omega_0\right)^2
    \vec A\left(x,y,z,\omega-\omega_0\right) \nonumber\\
\times \exp{\left(i\left(\omega-\omega_0\right)t\right)}d\omega.
\end{eqnarray}
The principe of causality request the next conditions on the
responce functions:

\begin{eqnarray}
\label {caus} \varepsilon (t-\tau)=0;\ \chi^{\left(3\right)}
\left(t-\tau_1,t-\tau_2,t-\tau_3\right)=0, \nonumber\\
 t-\tau<0;\ t-\tau_i<0;\ i=1,2,3.
\end {eqnarray}
That because we can prolonged the upper integral boundary to
infinity and to use standard Fourier presentation \cite{MN}:

\begin{eqnarray}
\label {intcs} \int\limits_{-\infty}^{t}
{\varepsilon_0(\tau-t)\exp{(i\omega\tau)}d\tau}=\int\limits_{-\infty}^{+\infty}
{\varepsilon_0(\tau-t)\exp{(i\omega\tau)}d\tau},\\
\int\limits_{-\infty}^{t}\int\limits_{-\infty}^{t}\int\limits_{-\infty}^{t}
\chi^{\left(3\right)} \left(t-\tau_1,t-\tau_2,t-\tau_3\right)
d\tau_1d\tau_2d\tau_3=\nonumber\\
\int\limits_{-\infty}^{+\infty}\int\limits_{-\infty}^{+\infty}\int\limits_{-\infty}^{+\infty}
\chi^{\left(3\right)} \left(t-\tau_1,t-\tau_2,t-\tau_3\right)
d\tau_1d\tau_2d\tau_3.
\end {eqnarray}

The spectral presentation of linear optical susceptibility
$\hat{\varepsilon_{0}}(\omega)$ is connected to the nonstationary
optical response function by the next Fourier transform:

\begin{eqnarray}
\label{a5} \hat{\varepsilon}_0(\omega)\exp{(-i\omega t)}
=\int\limits_{-\infty}^{+\infty}
{\varepsilon_0(t-\tau)\exp{(-i\omega\tau)}d\tau}.
\end{eqnarray}
Similar is the  expression for the spectral presentation of the
non-stationary nonlinear optical susceptibility
$\hat{\chi}^{(3)}$:

\begin{eqnarray}
\label{chi3} \hat{\chi}^{(3)}(\omega=2\omega-\omega)\exp{(-i\omega
t)}=
\int\limits_{-\infty}^{+\infty}\int\limits_{-\infty}^{+\infty}
\int\limits_{-\infty}^{+\infty}\chi^{\left(3\right)}
\left(t-\tau_1,t- \tau_2,t-\tau_3\right)\nonumber\\
\times\exp{\left(-i\left(\omega(\tau_1+\tau_2+\tau_3)\right)\right)}
d\tau_1d\tau_2d\tau_3.
\end{eqnarray}
From (\ref{eq6}), (\ref{a1}), (\ref{a2}) (\ref{intcs})and
(\ref{a5}) for the linear polarization of one of the complex
fields $\vec {P}_{lin}^{'}\left[\vec{A}\exp{(i(k_0z-\omega
t))}\right]$ is obtained:

\begin{eqnarray}
\label{a6}
 \vec {P}_{lin}^{'} =\int\limits_{\infty}^t \varepsilon _0
    \left(t-\tau\right)\exp{\left(i( k_0z-\omega_0\tau)\right)}
    \int\limits_{-\infty}^{+\infty}
    \vec A\left(x,y,z,\omega-\omega_0\right)\nonumber\\
\times \exp{\left(-i(\omega-\omega_0)\tau\right)}d\omega d\tau \nonumber\\
=\exp{(ik_0z)}\int\limits_{-\infty}^{+\infty}
    \vec A_i \left(x,y,z,\omega-\omega_0\right)
    \int\limits_{-\infty}^{+\infty}\varepsilon_0\left(t-\tau\right)
    \exp{(-i\omega \tau)} d\tau d\omega.
\end{eqnarray}
The second integral in (\ref{a6}) is simply equation (\ref{a5}),
and the linear polarization  can be written:

\begin{eqnarray}
\label{a7} \vec {P}_{lin}^{'}\left(x,y,z,t\right) =
\exp{\left(i(k_0z-\omega_0t)\right)}
 \int\limits_{-\infty}^{+\infty}
 \hat{\varepsilon}_0\left(\omega\right)
 \vec A\left(x,y,z,\omega-\omega_0\right)\nonumber\\
\times\exp{\left(-i\left(\omega-\omega_0\right)t\right)}d\omega.
\end{eqnarray}
It is important to point, that the similar expressions as
(\ref{a7}) are written also for the polarization of the complex
fields of kind:
\begin{eqnarray}
\label{rel1} \vec {P}_{lin}^{'}\left[\vec {A}\exp{(-i(k_0z-\omega
t))}\right]; \ \vec {P}_{lin}^{'}\left[\vec
{A}^*\exp{(i(k_0z-\omega t))}\right],
\end{eqnarray}
and
\begin{eqnarray}
\label{rel2}
 \vec {P}_{lin}^{'}\left[\vec {A}\exp{(i(k_0z-\omega t))}\right];\
 \vec {P}_{lin}^{'}\left[\vec {A}^*\exp{(-i(k_0z-\omega
 t))}\right].
\end{eqnarray}
The above procedures are used also to the nonlinear polarization.
When the third harmonic term is neglected the result is:
\begin{eqnarray}
\label{pnl2} \vec P^{nlin}(x,y,z,t)=\frac
{3}{4}\exp{\left(i(k_0z-\omega_0t)\right)}
\int\limits_{-\infty}^{+\infty}
\hat{\chi}^{(3)}(\omega)\exp{\left(-i(\omega-\omega_0)t\right)}\nonumber\\
\times|\vec A(x,y,z,\omega-\omega_0)|^2 \vec
A(x,y,z,\omega-\omega_0)d\omega.
\end{eqnarray}
Multiply (\ref{a7}), (\ref{pnl2}) by $1/c^2$, and using the second
derivatives in time, we obtain:

\begin{eqnarray}
\label{a8}
 \frac{1}{c^2}\frac{\partial^2\vec P_{lin}^{'}\left(x,y,z,t\right)}
{\partial t^2} =-\exp{\left(i\left( k_0z-\omega_0 t\right)\right)}
\int\limits_{-\infty}^{+\infty}
\frac{\omega^2\hat{\varepsilon}_0\left(\omega\right)}{c^2}
\vec A\left(x,y,z,\omega-\omega_0\right) \nonumber\\
\times \exp{\left(-i\left(\omega-\omega_0\right)t\right)}d\omega,
\end{eqnarray}
\begin{eqnarray}
\label{dpnl} \frac{4\pi}{c^2}\frac{\partial^2 \vec
P_{nlin}\left(x,y,z,t\right)}{\partial t^2} = -\exp{\left(i\left(
k_0z-\omega_0 t\right)\right)} \int\limits_{-\infty}^{+\infty}
\frac{3\pi\omega^2\hat{\chi}^{(3)}\left(\omega\right)}{c^2}\nonumber\\
\times|\vec A(x,y,z,\omega-\omega_0)|^2 \vec
A(x,y,z,\omega-\omega_0)
\exp\left(-i(\omega-\omega_0)t\right)d\omega.
\end{eqnarray}
The spectrum of the amplitude function is restricted, by writing
the wave vectors:
\begin{eqnarray}
k_{lin}=\sqrt{\frac{\omega^2\hat{\varepsilon}_0\left(\omega\right)}{c^2}},
\end{eqnarray}
and
\begin{eqnarray}
k_{nl}=\sqrt{\frac{3\pi\omega^2\hat{\chi}^{(3)}\left(\omega\right)}{c^2}}
\end{eqnarray}
near the carrying frequency in a Taylor series:

\begin{eqnarray}
\label{a9}
 k_{lin}^2\left(\omega\right)=
\frac{\omega ^2\hat{\varepsilon_0}\left(\omega\right)}{c^2} =
 k^{2}\left(\omega_0\right) +
\frac{\partial\left(k^2\left(\omega_0\right)\right)}{\partial\omega_0}
\left(\omega-\omega_0\right) \nonumber\\
+ \frac{1}{2}
\frac{\partial^2\left(k^2\left(\omega_0\right)\right)}{\partial
\omega_0^2} \left(\omega-\omega_0\right)^2 + ...,
\end{eqnarray}
\begin{eqnarray}
\label{nla9}
 k_{nl}^2\left(\omega\right)=
\frac{\omega ^2\hat{\chi}^{(3)}\left(\omega\right)}{c^2} =
k_{nl}^{2} \left(\omega_0\right) +
\frac{\partial\left(k_{nl}^2\left(\omega_0\right)\right)}{\partial\omega_0}
\left(\omega-\omega_0\right)+...
\end{eqnarray}
It is convenient to express the nonlinear wave vector by linear
wave vector and nonlinear refractive index:

\begin{eqnarray}
k_{nl}^2(\omega_0)=3\pi\omega_0^2\hat{\chi}^{(3)}(\omega_0)/c^2=
\frac{\omega_0^2\hat{\varepsilon}(\omega_0)}{c^2}\frac{3\pi\hat{\chi}^{(3)}(\omega_0)}
{\hat{\varepsilon}(\omega_0)} =k_0^2n_2,
\end{eqnarray}
where

\begin{eqnarray}
n_2(\omega_0)=\frac{3\pi\hat{\chi}^{(3)}(\omega_0)}{\hat{\varepsilon}(\omega_0)},
\end{eqnarray}
is the nonlinear refractive index. We note that any approximation
of the response function no used. There is only one requirement of
the spectral presentations (\ref{a5})and (\ref{chi3}) of the
response functions: to admit first and second order derivatives in
respect to frequency (to be smooth functions). So, this method is
not limited from the time duration of the response function. This
give us the possibility to apply it also when the half-max of the
pulses are in order of the time duration of the response function
(femtosecond pulses). The restriction is only in respect of the
relation between the main frequency $\omega_0$ and time duration
of the envelope functions $t_0$ determinate from the relations
(\ref{a9}) and (\ref{nla9}) (conditions for slowly varying
amplitudes). Putting eqn.(\ref{a9})and (\ref{nla9}) in
(\ref{a8})and (\ref{dpnl}) respectively and keeping in mind the
expressions of time derivative of the amplitude functions
(\ref{a3})-(\ref{a4}) for the electric field, the second time
derivative of the linear polarization of the optical component is
represented in next truncated form:

\begin{eqnarray}
\label{a10}
 \frac{1}{c^2}\frac{\partial^2\vec P_{lin}^{'}\left(x,y,z,t\right)}{\partial t^2}
=\left(-k_0^2 \vec A +\frac{2ik_0}{v}\frac{\partial \vec
A}{\partial t} + \left(k_0 k_0^{"}+\frac{1}{v^2}\right)
\frac{\partial^2\vec A}{\partial t^2}+... \right)\nonumber\\
\times \exp{\left(i(k_0z-\omega_0t)\right)}.
\end{eqnarray}
Similar result is obtained also for the linear polarization of the
terms of kind (\ref{rel1}) and(\ref{rel2}). The nonlinear
polarization therm is:
\begin{eqnarray}
  \label{nla10}
 \frac{4\pi}{c^2}\frac{\partial^2\vec P_{nlin}\left(x,y,z,t\right)}{\partial t^2}
= \left(-k_0^2 n_2\left|\vec A\right|^2\vec A
+2ik_{nl}\frac{\partial
k_{nl}}{\partial\omega}\frac{\partial\left(\left|\vec
A\right|^2\vec A\right)}{\partial t}
+ ... \right)\nonumber\\
\times \exp{\left(i(k_0z-\omega_0t)\right)}=\nonumber\\
\left(-k_0^2 n_2\left|\vec A\right|^2\vec A
+\left(2ik_{0}\frac{n_2}{v}+ik_0^2\frac{\partial
n_2}{\partial\omega}\right)\frac{\partial\left(\left|\vec
A\right|^2\vec A\right)}{\partial t}
+ ... \right)\nonumber\\
\times \exp{\left(i(k_0z-\omega_0t)\right)},
\end{eqnarray}
where $v=1/(\partial k_{lin}\left(\omega_0\right)/\partial\omega)$
and $k_0^{"} $ are the group velocity, and dispersion. From the
wave equation (\ref{eq9}), using (\ref{xx1}), (\ref{a10}),
(\ref{nla10}) and (\ref{rel1}),(\ref{rel2}), the next two
equations of complex vector amplitude and it complex conjugate are
obtained:

\begin{eqnarray}
\label{a11} i\left(\frac{\partial\vec A}{\partial t} +
v\frac{\partial\vec A}{\partial z}+\left(n_2+\frac {
k_0v}{2}\frac{\partial n_2}{\partial\omega}\right)\frac{\partial
\left(\left|\vec A\right|^2\vec
A\right)}{\partial t} \right) =\nonumber\\
\frac{v}{2k_0}\Delta\vec A -
\frac{v}{2}\left(k_0^{"}+\frac{1}{k_0v^2}\right)
\frac{\partial^2\vec A}{\partial t^2} + \frac{n_2 k_0
v}{2}\left|\vec A\right|^2\vec A,
\end{eqnarray}
\begin{eqnarray}
\label{a11c} -i\left(\frac{\partial\vec A^*}{\partial t} +
v\frac{\partial\vec A^*}{\partial z}+\left(n_2+\frac {
k_0v}{2}\frac{\partial n_2}{\partial\omega}\right)\frac{\partial
\left(\left|\vec A\right|^2\vec
A^*\right)}{\partial t} \right)=\nonumber\\
\frac{v}{2k_0}\Delta\vec A^* -
\frac{v}{2}\left(k_0^{"}+\frac{1}{k_0v^2}\right)
\frac{\partial^2\vec A^*}{\partial t^2} + \frac{n_2 k_0
v}{2}\left|\vec A\right|^2\vec A^*=0.
\end{eqnarray}

The equations (\ref{a11}) and (\ref{a11c}) are equal and solving
(\ref{a11}) for $\vec A$ we can find also $\vec A^*$ and
respectively the real amplitude vector field $\vec A'$.  We
investigate the case when:
\begin{eqnarray}
\label {a12}
v^2k_0k_0^{"} = - 1,
\end{eqnarray}
a condition that can only be fulfilled for materials possessing
negative dispersion.The $\frac{\partial^2\vec A}{\partial t^2}$
term in this case is neglected. Applying a Galilean transformation
to the vector amplitude equation (\ref{a11}), where the new
reference frame moves at the group velocity, $t' = t; z' = z - vt$
(the primes are missed for clarify) , we obtain our final vector
amplitude equation:

\begin{eqnarray}
\label {a13} -i\left(\frac{\partial\vec A}{\partial t}+
\left(n_2+\frac { k_0v}{2}\frac{\partial
n_2}{\partial\omega}\right)\frac{\partial \left(\left|\vec
A\right|^2\vec A\right)}{\partial t} \right)+\nonumber\\
\frac{v}{2k_0}\Delta_{\bot}\vec A -
\frac{v^3k_0^{"}}{2}\frac{\partial^2\vec A}{\partial z^2} +
\frac{n_2 k_0 v}{2}\left|\vec A\right|^2\vec A=0,
\end{eqnarray}
where $\Delta_{\bot}=\frac{\partial^2}{\partial x^2} +
\frac{\partial ^2}{\partial y^2}$.

\section {Norming VNSE. Exact vortex solutions}

In this section we will obtain solutions in the case of amplitude
vector equations of one carrying frequency. The case of two
carrying frequencies is discussed in \cite {AK}. The case of three
and more localized optical waves on different frequencies with
additional conditions on the frequencies (parametric vortices)
will be discussed in a next paper. Defining the rescaled
variables:

\begin{eqnarray}
\label{eq12} \vec A=A_0\vec A';\ x=r_0x';\ y=r_oy';\  z=r_0z';\
t=t_0t',
\end{eqnarray}
 and constants:

\begin{eqnarray}
\label{norm}
 \alpha=2k_0r_0^2/t_0v;\  \beta=v^2k"k_0;\
\gamma=k_0^2r_0^2n_2\left|A_0\right|^2;\nonumber\\
\gamma_1=k_0r_0n_2\left|A_0\right|^2;\ \gamma_2=\frac {v}{c}
k_0r_0n_2\left|A_0\right|^2,
\end{eqnarray}
 equation (\ref{a13}) can be transformed in the following (the primes are
 not written):
\begin{eqnarray}
\label {eq13}
 -i\left(\alpha\frac{\partial\vec A}{\partial t}+
(\gamma_1+\gamma_2)\frac{\partial \left(\left|\vec A\right|^2\vec
A\right)}{\partial t} \right)+\nonumber\\
\Delta_{\bot}\vec A -\beta\frac{\partial^2\vec A}{\partial z^2} +
\gamma\left|\vec A\right|^2\vec A=0.
\end{eqnarray}
The dispersion therm (the second derivative in 'z' direction of
the amplitude function) in transparency region of a media is
usually one or two order of magnitude smaller than the diffractive
therm (transverse Laplasian). That because usually  the linear
(dispersion) parameter is  $\beta\sim 10^{-2}$. There is
possibility to reach $\beta=-1$ only in some special cases, near
to Lengmuire frequency in cold plasma, high-frequency transparency
region of dielectrics and near to some of the electronic
resonances \cite {AK}. The constant $\alpha$ has a value of
$\alpha\approx 10^2$ ($\alpha \approx r_0 k_0 $) if the of slowly
varying approximation is used. When the nonlinear constant admit
typical, critical for self-focusing regime value $\gamma=1$, the
constants $\gamma_1\cong\gamma_2\sim10^{-2}$ are small. We point
here that the effects of asymmetry of the pulses (asymmetry of
their spectrum) due to nonlinear addition to the group velocity
presented by second therm in equation (\ref{eq13}) is substantial
when $\gamma_1\cong\gamma_2\sim 1$ or $\gamma\sim10^{2}$. These
terms depends from intensity of the fields and when
$\gamma\sim10^{2}$ or $n_2\left|A_0\right|^2\sim 10^{-2}$ effects
of self-steepening of the pulses, connected also with considerable
self-focusing can be estimated. Such type of experiments
\cite{FORK} are provided with $80$ femtosecond pulses and
intensity $I_0\sim 10^{14} Wt/cm^2$. This correspond to the
parameters discussed above ($\gamma\sim10^{2}$ or
$n_2\left|A_0\right|^2\sim 10^{-2}$). In this paper is
investigated the case, when $\beta=-1$ (negative dispersion) and
$\gamma=1$. Neglecting the small terms the equation (\ref{eq13})
becomes:

\begin{eqnarray}
\label{eq15}
-i\alpha\frac{\partial \vec A}{\partial t} + \Delta_{\bot}\vec A +
\frac{\partial^2\vec A}{\partial z^2} +
\left|\vec A\right|^2\vec A = 0.
\end{eqnarray}
Now we going to the next step - the possible polarization of the
amplitude vector field (\ref{eq15}). Recently, this problem was
discussed largely \cite{BORN,BERG} and was pointed the different
between the polarization of the optical waves with spectral
bandwidth (slowly varying amplitudes), and the polarization of
monochromatic fields. In case of monochromatic and
quasi-monochromatic fields the Stokes parameters can be
constructed from transverse components of the wave field
\cite{BORN}. This lead to two component vector fields in a plane,
transverse to direction of propagation. For electromagnetic fields
with spectral bandwidth (our case) the two dimensional coherency
tensor cannot be used and the Stokes parameters cannot be found
directly. As it was shown by T.Carozzi and all in \cite{BERG},
using high order of symmetry (SU(3)), six independent Stokes
parameters can be found. This corresponds to three component
vector field. Here is investigate this case. The increasing of the
spectral bandwidth of the vector wave, increasing also the
depolarization term (component, normal to the standard Stokes
coherent polarization plane). The amplitude vector function of
electrical field $\vec{A}$ is represented as sum of three
orthogonal linearly polarized amplitudes:

\begin{eqnarray}
\label{eq16} \vec A(x,y,z,t)=\sum\limits_{\vec j= \vec x, \vec y,
\vec z} {\vec jA_j(x,y,z)\exp {\left(i\Omega t \right)}}.
\end{eqnarray}
In a Cartesian coordinate system, for solutions of the kind of
(\ref{eq16}), the vector equation is reduced to a scalar system of
three nonlinear wave equations:

\begin{eqnarray}
\label{eq17} \alpha\Omega A^{l} + \Delta A^{l} + \sum\limits_{j =
x,y,z}{\left(\left|A^j\right|^2\right) A^l} = 0;\ l = x,y,z.
\end{eqnarray}
As it was pointed in \cite {MN}, we may choose to express each of
the components $A_i$ in spherical coordinates
$A_i(r,\theta,\varphi)$ of the independent variables $i=x,y,z$. In
this way  the linear unique (polarization) vector of each of the
components hold down. The system (\ref{eq17}) in spherical
variables is:

\begin{eqnarray}
\label{eq18} \alpha\Omega A^l + \Delta_r A^l +
\frac{1}{r^2}\Delta_{\theta ,\varphi}A^l + \sum\limits_{j =
x,y,z}{\left(\left|A^j\right|^2\right) A^l} = 0;\ l = x,y,z,
\end{eqnarray}
where

\begin{eqnarray}
\label{eq19}
\Delta_r=
\frac{1}{r^2}\frac{\partial}{\partial r}\left(r^2\frac{\partial}{\partial r}
\right);
\end{eqnarray}

\begin{eqnarray}
\label{eq20}
\Delta_{\theta ,\varphi}=
\frac{1}{\sin \theta}\frac{\partial}{\partial \theta}
\left(\sin\theta \frac{\partial}{\partial \theta}\right) +
\frac{1}{\sin^2\theta}\frac{\partial^2}{\partial \varphi^2},
\end{eqnarray}
are the radial and the angular operator respectively and
  $r=\sqrt{x^2 + y^2 + z^2}$
  $\theta= \arccos\frac zr$
  $\varphi=arctan\frac xy $,
are the moving spherical variables of the independent variables
$x, y, z$. The system of equations (\ref{eq18}) are solved, using
the method of separation of the variables. We present the
components of the field as a product of a radial and an angular
part:

\begin{eqnarray}
\label{eq21}
A^l(r,\theta ,\varphi) = R(r)Y_l(\theta ,\varphi);
l = x,y,z,
\end{eqnarray}
 with the additional constraint on the angular parts:

\begin{eqnarray}
\label{eq22}
{\left|Y_x\left(\theta, \varphi\right)\right|}^{2} +
{\left|Y_y\left(\theta, \varphi\right)\right|}^{2} +
{\left|Y_z\left(\theta, \varphi\right)\right|}^{2} = const.
\end{eqnarray}
Multiplying equation (\ref{eq18}) by $\frac{r^2}{RY_l}$, and
bearing in mind the constraint expressed in (\ref{eq22}), we
obtain:

\begin{eqnarray}
\label{eq23} \frac{r^2\Delta_rR}{R} + r^2\left(\alpha\Omega  +
{\left|R\right|}^2\right) =
-\frac{\Delta_{\theta,\varphi}Y_l}{Y_l}=\ell(\ell + 1),
\end{eqnarray}
where $\ell$ is a number. Thus the following equations for the
radial and the angular parts of the wave functions are obtained:

\begin{eqnarray}
\label{eq24} \Delta _r R + \alpha\Omega R + {\left|R\right|}^2R -
{\frac{\ell(\ell + 1)}{r^2}}R = 0
\end{eqnarray}

\begin{eqnarray}
\label{eq25} \Delta_{\theta,\varphi}Y_l +\ell(\ell+1)Y_l=0;\ l =
x,y,z.
\end{eqnarray}
Equations (\ref{eq24})-(\ref{eq25}) shows that the nonlinear term
occurs only in the radial components of the fields. As for the
angular parts we have the usual linear eigenvalue problem. The
solutions of equations (\ref{eq25}) of the angular parts are well
known; each of them has exact solutions of the form:

\begin{eqnarray}
\label{eq26}
Y_l=Y_{\ell}^{m}=\Theta_{\ell}^{m}(\theta)\Phi_{m}(\varphi ) =
\sqrt {\frac{4\pi}{3}}
\sqrt {\frac{2\ell+(\ell-m)!}{4\pi(\ell + m)!}}
P_{\ell}^{m}(\cos \theta)\exp \left( im\varphi \right),
\end{eqnarray}
where $P_{\ell} ^{m}$ are the Legendre's functions for a discrete
series of numbers: $\ell = 0,1,2....;m = 0,\pm 1,\pm 2,..$ and
${\left| {m} \right|} < \ell $. Returning to equation
(\ref{eq18}), it is seen that separation of variables for
spherical functions, which satisfy condition (\ref{eq22}), is
possible only for $l=1$:

\begin{eqnarray}
\label{eq27}
 Y_x =Y_1^{ - 1} = \sin \theta \cos \varphi \nonumber\\
 Y_y =Y_1^1 = \sin \theta c\sin \varphi  \nonumber\\
 Y_z = Y_1^0 = \cos \theta,
\end{eqnarray}
or another appropriate combination of axes. By choosing one of
these angular components for each of the field components we see
that the eigenfunctions (\ref{eq27}) are solutions to the angular
part of the set of equations (\ref{eq25}). It is straightforward
to show that the radial part of the equations (\ref{eq24}) admit
``de Broglie soliton'' solutions \cite {BAR} of the form:

\begin{eqnarray}
\label{eq28} R =\sqrt 2\frac{\exp{\left(i\sqrt{\alpha\Omega}
r\right)}}{r}.
\end{eqnarray}
In this way, we prove the existence of vortex solutions with
angular momentum $l = 1$ in a Kerr type medium. The solutions of
the vector amplitude equation (\ref{eq15}) in a fixed basis are:

\begin{eqnarray}
\label{c1} A_x = \sqrt 2\frac{\exp\left({i\sqrt{\alpha\Omega}
r}\right)}{r} \sin \theta \cos \varphi \exp \left(i\Omega t
\right),
\end{eqnarray}

\begin{eqnarray}
\label{c2} A_y = \sqrt 2\frac{\exp\left({i\sqrt{\alpha\Omega}
r}\right)}{r} \sin \theta \sin \varphi \exp \left(i\Omega t
\right),
\end{eqnarray}

\begin{eqnarray}
\label{c3} A_z = \sqrt 2\frac{\exp\left({i\sqrt{\alpha\Omega}
r}\right)}{r} \cos \theta \exp \left(i\Omega t \right).
\end{eqnarray}
The equation (\ref{eq15}) is normed and this lead to the next
normed constants:

\begin{eqnarray}
\alpha\Omega=1;\ \Omega=\sim10^{-2}.
\end{eqnarray}
Now we can turn bask to the amplitude of real solutions $\vec
{A}^{'}$, using equation (\ref{qq1}):

\begin{eqnarray}
\label{eq30} A_x^{'}=\frac{1}{2i}\left( A_x-A_x^{*}\right)=\nonumber\\
\sqrt{2}\frac{\sin\left(\sqrt{\alpha\Omega}r+\Omega
t\right)}{r}\sin \theta \cos \varphi ,
\end{eqnarray}

\begin{eqnarray}
\label{eq31} A_y^{'}=\frac{1}{2i}\left( A_y-A_y^*\right)=\nonumber\\
\sqrt {2} \frac{\sin\left({\sqrt{\alpha\Omega} r+\Omega
t}\right)}{r}\sin \theta \sin \varphi ,
\end{eqnarray}
\begin{eqnarray}
\label{eq32} A_z^{'}=\frac{1}{2i}\left( A_z-A_z^*\right)=\nonumber\\
\sqrt {2} \frac{\sin\left({\sqrt{\alpha\Omega} r+\Omega
t}\right)}{r}\cos \theta.
\end{eqnarray}
The condition $\nabla \cdot\vec {E} = 0$ separates the nonlinear
Maxwell's system in two wave equation system, one nonlinear, of
the electric field, and another nonlinear wave equation of the
magnetic field. We define again a complex presentation of the real
magnet field by the relation:

\begin{eqnarray}
\label{h1}
 \vec H\left(x,y,z,t\right)=
\vec {C}'\left(x,y,z,t\right)
\frac{1}{2i}\left(\exp{\left(i(k_0z-\omega_0t)\right)}-c.c\right),
\end{eqnarray}
where $\vec C' $ and $\omega_0$ and $k_0 $ are the real vector
amplitude, the optical frequency and the wave number of the
optical field respectively. The real magnet vector amplitude $\vec
C' $ is represented also by a complex vector amplitude field:

\begin{eqnarray}
\label{h2}
 \vec C'\left(x,y,z,t\right)=
\frac{1}{2i}\left(\vec C\left(x,y,z,t\right)-\vec
C^*\left(x,y,z,t\right)\right),
\end{eqnarray}
where $\vec C$ is the complex amplitude of the magnet field.
Applying similar procedures to those who were made for electric,
namely; using the Fourier representation of the amplitude
functions $\vec C$ and their time derivatives to second order, as
were done for the electric, we obtain the vector equation of
slowly varying amplitudes of the magnetic field $\vec {C}$:

\begin{eqnarray}
\label{eq35} i\alpha\frac{\partial\vec C}{\partial t} +
\Delta_{\bot}\vec C +\frac{\partial^{2}\vec C}{\partial z^{2}} +
{\left|\vec A\right|}^2\vec C= 0,
\end{eqnarray}
if the condition:

\begin{eqnarray}
\label{eq36}
\nabla {\left| {\vec {A}} \right|}^{2}\times \vec {A} = 0,
\end{eqnarray}
is satisfied. For solutions of the amplitudes of electrical field
like (\ref{c1})-(\ref{c3}), the condition (\ref{eq36}) is
satisfied. Equation (\ref{eq35}) has a localized solution of the
same kind (\ref{c1})-(\ref{c3}), as the amplitude functions of the
electric field, but with opposite phase:

\begin{eqnarray}
\label{eq37} C_x = \sqrt 2\frac{\exp\left({i\sqrt{\alpha\Omega}
r}\right)}{r} \sin \theta \cos \varphi \exp \left(-i\Omega t
\right),
\end{eqnarray}

\begin{eqnarray}
\label{eq38} C_y = \sqrt 2\frac{\exp\left({i\sqrt{\alpha\Omega}
r}\right)}{r} \sin \theta \sin \varphi \exp \left(-i\Omega t
\right),
\end{eqnarray}

\begin{eqnarray}
\label{eq39} C_z = \sqrt 2\frac{\exp\left({i\sqrt{\alpha\Omega}
r}\right)}{r} \cos \theta \exp \left(-i\Omega t\right).
\end{eqnarray}

The real dynamics of localized fields can be understood only by
investigating both equation (\ref{eq13}) and (\ref{eq35}).The
solutions of electrical (\ref{c1})- (\ref{c3}) and magnet
(\ref{eq37}) - (\ref{eq39}) fields are with opposite phase, and
that because the corresponding electromagnetic localized wave
oscillating in coordinate system moving with group velocity.

\section{Experimental conditions}

     The results presented above require the conditions relating the linear
parameter $\beta $ and the nonlinear parameter $\gamma $ to
satisfy: namely $\beta=-1$ and $\gamma = 1$. The relation
$\gamma=1$ is a typical critical value for starting of
self-focusing regime of a Gausian pulse. The nonlinear parameter
written again is:

\begin{eqnarray}
\label{eq42}
\gamma = r_{0}^{2} k_{0}^{2} n_{2} {\left| {A_{0}}  \right|}^{2} = \alpha
^{2}n_{2} {\left| {A_{0}}  \right|}^{2} = 1.
\end{eqnarray}
The constant $\alpha = r_{0}k_{i}$ in the optical region ranges in
value from:

\begin{eqnarray}
\label{eq43}
\alpha \cong 10^{2} - 10^{3}.
\end{eqnarray}
Using this range in the condition (\ref{eq42}), we obtain a
required nonlinear refractive index change:

\begin{eqnarray}
\label{eq44}
n_{2} {\left| {A_{0}}  \right|}^{2} \cong 10^{ - 4} - 10^{ - 6}.
\end{eqnarray}
It was shown, using relations (\ref{norm}), that if we investigate
near to the critical value of self-focusing $\gamma=1$, then
$\gamma_1\sim\gamma_2\sim10^{-2}-10^{-3}$. This is one important
result: The nonlinear addition to the group-velocity therm in
amplitude equation (\ref{eq13}) is relatively small, even for
(femtosecond) pulses with time duration equal or little to
characteristic time response of the nonlinear media.

Using the condition of linear parameter $\beta$:

\begin{eqnarray}
\label{eq45}
\beta = k_{0}v^{2}k" = - 1,
\end{eqnarray}
we see that the vortex propagation take place only in the
negative-dispersion region. In \cite{AK} we find, that the
constraint, given in equation (\ref{eq45}) correspond to the next
two experimental situations:

 1. Cold plasma near the Langmuir frequency.

 2. A region near an electronic resonance in an isotropic medium.

There is also a equivalence in high-frequency region between the
spectral presentation of dielectric susceptibility of a cold
plasma and this of a dielectric media. The expression is equal
wish precise a (dielectric) constant. Near these frequencies the
dispersion parameter increases rapidly and admits values:

 $\mid k"\mid\sim 10^{-24}-10^{-25} \frac{cek^2}{cm}$
This lead to the fact, that the dispersion term (the second
derivative in z direction of the amplitude function) in normed
amplitude equations of electric (\ref{eq13}) and magnet
(\ref{eq35}) field is of the same order of the diffractive term
(transverse Laplasian).

\section {Finiteness of the energy for the vortex solutions}

 Now we come to the crucial point - energy and shape of the vortex solutions
 (\ref{eq30})-(\ref{eq32}) and (\ref{eq37})-(\ref{eq39}) of the electric and
 magnet fields. To prove the finiteness of energy of the vortex solutions
 (\ref{eq30})-\ref{eq32}) and (\ref{eq37})-(\ref{eq39}),
 we start with the equations for averaged in time balance of energy density
 of electrical and magnet field \cite{LAN}:

\begin{eqnarray}
\label{b3}
\frac{\partial W}{\partial t} =
\frac {1}{16\pi}\left(\vec E\cdot\frac{\partial \vec D^{*}}{\partial t} +
\vec E^{*}\cdot\frac{\partial \vec D}{\partial t} +
  \vec B\cdot\frac{\partial \vec B^{*}}{\partial t} +
  \vec B^{*}\cdot\frac{\partial \vec B}{\partial t}\right),
\end{eqnarray}
where $\vec D=\vec P_{lin}+4\pi\vec P_{nlin}$ is a sum of the
linear induction and the nonlinear induction of the electrical
field. In  more of the books the calculations of the averaged
energy of the optical waves in a dispersive media are worked out
bearing in mind only the first order of slowly varying amplitude
approximation of the linear electrical induction. That is because
the result comes to the old result of Brillouin 1921 for energy
density of electrical field:

\begin{eqnarray}
\label{b4}
W_{lin} =
\frac {1}{8\pi}\left(\frac{\partial\left(\omega\epsilon_0\right)}
{\partial \omega}{\left|\vec A\right|}^2+ {\left|\vec C\right|}^2\right)
\nonumber\\
=\frac {1}{8\pi}\left(\frac{c}{v}
{\left|\vec A\right|}^2+ {\left|\vec C\right|}^2\right).
\end{eqnarray}
It is straightforward to show using this approximation that after
integrating in space our vortex solutions they admit infinite
energy. As seen from (\ref{b4}), this approximation do not include
the dispersion parameter $k"$, which is present in the envelope
equations (\ref{eq13}), (\ref{eq35}) as coefficient in front of
the second derivative in z direction. So there is no possibility
to estimate the influence of $k"$ in the energy integral. Our
vector amplitude equations are obtained, using the second order
approximation of $\vec P_{lin}$. The right expression of the
energy density in this case requires to expand the linear
electrical inductions in the energy integral also up to the second
order, and to include this dispersion parameter in the energy
integral. Using truncated form for first derivative in time of the
linear induction of electrical field we have:

\begin{eqnarray}
\label{b5} \frac{\partial{\vec P_{lin}}}{\partial
t}=-i\omega_0\epsilon(\omega_0)\vec A+ \frac
{c}{v}\frac{\partial{\vec A}}{\partial t}+ \frac
{ick^"}{2}\frac{\partial^2{\vec A}}{\partial t^2}.
\end{eqnarray}
From (\ref{b3}), (\ref{b5}) and complex conjugate of (\ref{b5}) we obtain:

\begin{eqnarray}
\label{b6} \frac{\partial W_{lin}}{\partial t} = \frac
{c}{v}\left(\vec A\frac{\partial \vec A^{*}}{\partial t} + \vec
A^{*}\frac{\partial \vec A}{\partial t}\right)+ \frac{ic
k^"}{2}\frac{\partial}{\partial t} \left(\vec A^{*}\frac{\partial
\vec A}{\partial t}- \vec A\frac{\partial \vec
A^{*}}{\partial t}\right)\nonumber\\
+ \vec B\frac{\partial \vec B^{*}}{\partial t} +
\vec B^{*}\frac{\partial \vec B}{\partial t},
\end{eqnarray}
 or
\begin{eqnarray}
\label{b7} W_{lin} = \frac {c}{v}{\left|\vec A\right|}^2+
{\left|\vec C\right|}^2 +\frac {ic k^"}{2}\left(\vec
A^{*}\frac{\partial \vec A}{\partial t} -
  \vec A\frac{\partial \vec A^{*}}{\partial t}\right),
\end{eqnarray}
where $A^{*}$ denote complex conjugate in time amplitude function.
Rewriting the vortex solutions of electrical
(\ref{eq30})-(\ref{eq32}) and magnet (\ref{eq37})-(\ref{eq39})
fields in not normed (dimension) coordinates and substituting in
(\ref{b7}), for the case of negative dispersion finally we obtain
the next expression of the linear part of averaged energy density:

\begin{eqnarray}
\label{b8} W_{lin} = \left(1+\frac {c}{v}-c\mid
k^"\mid\triangle\omega\right){\vec {A'}}^2,
\end{eqnarray}
where $\triangle\omega$ denote the spectral bandwidth of the
vortices. We obtain one unexpected result: with the increasing of
the spectral bandwidth of our solutions the linear part of energy
density will decrease in the negative dispersion region.
Conditions when the linear part of the energy density of
electrical field is zero determine one critical spectral bandwidth
$\triangle\omega_c$:

\begin{eqnarray}
\label{b9} \triangle\omega_c=\frac {v+c}{vc}\frac{1}{\mid k"\mid}.
\end{eqnarray}
Near to plasma frequency or some of the electronic resonances the
dispersion parameter increases and values about: $k"\sim
10^{-24}-10^{-25} \frac{cek^2}{cm}$. The critical spectral
bandwidth in this case becomes: $\triangle\omega_c\sim
10^{14}-10^{15}$ Hz. The envelope approximation request, that:
$\omega_0>\triangle\omega_c$. This condition shows that the linear
part of energy density of vortices is zero  only when their main
frequency is situated in optical region, or regions, which
frequency is  greater than the optical and admit spectral
bandwidth equal to $\triangle\omega_c$. The nonlinear part of
averaged energy density is expressed in \cite{SHEN} and for the
vortex solutions (\ref{c1})-(\ref{c3}) it becomes:

\begin{eqnarray}
\label{b10} W_{nlin} = n_2{\left|\vec {A'}\right|}^2\vec {A'}^{2}+
 \omega_0\frac{\partial n_2}{\partial\omega_0}.
 {\left|\vec {A'}\right|}^2\vec{A'}^2.
\end{eqnarray}
These results give the conditions for the finiteness of energy of
the vortices: the spectral bandwidth of vortices to be equal to
$\triangle\omega_c$. Integrating $W_{nlin}$ on the 3D space we
obtained a finite value proportional to the main frequency
$\omega_0$. The intensity part in this expression for the energy
(\ref{b10}) is limited by the experimental condition (\ref{eq44})
and it is a constant. In this way we obtain the result, that the
localized energy of the solutions increasing linearly with the
increasing of the main frequency. The numerical investigation of
stability for such type vortices, using split-step Fourier method,
is provided in \cite{AK}. Comparing the numerical calculations for
vortices with these, of a standard Gausian pulse, it can be
clearly seen that the vortices propagate without any changing of
their shape, whereas the Gaussian pulse, with the same initial
amplitude, self-focuses rapidly in a much shorter distance.

      In additional, we also investigate the shape and the behavior of
 electrical and magnet field in the origin and infinity. To obtain the
 correct result, the real part solutions of the amplitude equations
 (\ref{eq31})-(\ref{eq32}) and (\ref{eq37})- (\ref{eq39}) must be rewritten
 in the independent Cartesian coordinates x, y, z:

\begin{eqnarray}
\label {s1}
A'_x=\frac{x\sin(\sqrt{\alpha\Omega}\sqrt{x^2+y^2+z^2}+\Omega t)}
{x^2+y^2+z^2},
\end{eqnarray}
\begin{eqnarray}
\label {s2}
A'_y=\frac{y\sin(\sqrt{\alpha\Omega}\sqrt{x^2+y^2+z^2}+\Omega t)}
{x^2+y^2+z^2},
\end{eqnarray}
\begin{eqnarray}
\label {s3}
A'_z=\frac{z\sin(\sqrt{\alpha\Omega}\sqrt{x^2+y^2+z^2}+\Omega t)}
{x^2+y^2+z^2}.
\end{eqnarray}
The limits of these functions in origin and in infinity are
investigated, using the Haine criteria for limit of a
multidimensional function. It is straightforward to show that the
solutions (\ref{s1})-(\ref{s3}) are odd functions, going to zero
in infinity and admit very little value near to the origin. In the
origin, using the same criteria, we find that the above functions
have no limit. But as we must point out, the envelope function
(\ref{s1})-(\ref{s3}) is only imagine functions around the real
electric field, expressed in our case by the relation (\ref{a1}):

\begin{eqnarray}
\label{b2}
E_x=A'_x\sin{(k_0z-\omega_0t)}\\
E_y=A'_y\sin{(k_0z-\omega_0t)}\\
E_z=A'_z\sin{(k_0z)-\omega_0t}.
\end{eqnarray}
Using the same criteria as in the previous case, it is seen, that
the electrical field admit exact limit zero in the origin and
infinity:
\begin{eqnarray}
lim_{x,y,z\rightarrow 0}(E_i)=0\nonumber\\
lim_{x,y,z\rightarrow \pm\infty}(E_i)=0;\ i=x,y,z.
\end{eqnarray}

\section{Conclusion}
The applied method of slowly varying amplitudes of the electrical
and magnet vector fields give us the possibility to reduce the
nonlinear vector integro-differential equations  to vector
nonlinear differential equations of amplitudes. Here is the place
to explain more some of the differences between the solutions,
obtained by separation of the variables of the usual linear scalar
Schredinger equation with potential depending only on 'r' (for
example hydrogen atom) and solutions of the vector version of NSE.
For the linear Schredinger equations with potential, to higher
order spherical functions correspond more higher order of radial
spherical Bessel functions. While for vector version of NSE to
higher order of spherical functions ($\ell=1,2,..$) correspond
higher number of the fields components and higher value of
localized energy. We have for all radial solutions the zero
spherical Bessel function $\frac{\sin\alpha r}{r}$ in nonlinear
case. This is the reason to start with such special kind of
complexification of our electrical field (\ref{a1}). It is
important to highlight also the equivalence in high frequency
region between the linear dielectric susceptibility of a cold
plasma and this of a dielectric media. A plasma frequency of
dielectric media in this region can determine also. The expression
is equal to this of cold plasma wish precise a (dielectric)
constant. This discussion shows two possible regions for observing
optical vortices:
  $1.$ High frequency (transparency) region of cubic dielectrics or cold plasma.
  $2.$ Near to some of the electronics resonances of   $\chi^{(3)}$
  media.
Above mentioned investigations are provided only for the real part
of linear and nonlinear susceptibility. That is why the first
possibility, high frequency region, is more attractive for
observing of optical vortices. The complex part of linear
susceptibility is significant near to electronic resonances and
should be get in mind in a more detailed analysis.


\newpage
\begin{thebibliography}{100}

\bibitem {KEL}
 P.L.Kelley, "Self-focusing of optical beams", Phys. Rev. Lett. 15,
 1005-1008(1965).
\bibitem {KARP}
 V.I.Karpman, \textit{Nonlinear Waves in Dispersive Media},
  (Nauka, Moskow, 1973).
\bibitem{MN}
 J.V.Moloney and A.C.Newell, \textit{Nonlinear Optics}, (Addison-Wesley Publ.
  Comp.,1991).
\bibitem {JMN}
  J.A.Powell, J.V.Moloney and A.C.Newell, JOSA B, 10, N7, 1230-1241 (1993).
\bibitem {AK}
  D.R.Andersen and L.M.Kovachev, JOSA B, 19, N3, 376-384 (2002).
\bibitem {SIL}
 Y.Silberberg, Opt.Lett. 15,1282(1990).
\bibitem {EE}
  D.E. Edmundson and R.H.Enns, Opt.Lett. 18, 1609-1611 (1993).
\bibitem{MWBL}
 R.Mc.Leod, K.Wagner, and S.Blair, Phys.Rev.A, 52, 3254-3278 (1995).
\bibitem {AHM}
 J.M. Soto-Crespo,E.M. Wright and N.N.Akhmediev, Phys.Rev.A, 45, 3168 (1992).
\bibitem {EDE}
 D.E.Edmundson and R.H.Enns, Phys.Rev. A, 51, 2491-2497 (1995).
\bibitem {SW}
 G.A.Swartzlander, Jr., C.T.Law, Phys. Rev. Lett., 69, 2503 (1992).
\bibitem {CRIS}
  J.Christou, V.Tikhonenko, Yu.S.Kivshar, B.Luter-Daves, Opt.Lett., 21, 1649-
  1651 (1996).
\bibitem {STA}
  K.Stalinas, Chaos Solitons Fractals, 4, 1783-1796.
\bibitem {CHI}
 E.M.Wright, R.Y.Chiao and J.C.Garrison, Chaos Solitons Fractals,
 4, 1783-1796.
\bibitem {MAM}
 A.V.Mamaev, M.Soffman and A.A.Zozula, Phys.Rev.Lett., 76, 2262-2265 (1996).
\bibitem{TCRIS}
 V.Tikhonenko, J.Christou, B.Luter-Daves, JOSA B, 12, 2046-2052 (1995).
\bibitem {TOR}
 L.Torner,D.V.Petrov, Electron. Lett., 33, 608 (1997).
\bibitem {SKR}
 W.J.Firth, D.V.Skryabin,  Phys.Rev.Lett., 77, 2450 (1997).
\bibitem {DESM}
 A.Desyatnikov A.Maimistov and B.Malomed Phys. Rev. E, 61, 3107-3113 (2000).
\bibitem {MAL}
 B.A.Malomed, L.C.Grasovan, D.Mihalache, Physica D, 161, 187-201 (2002).
\bibitem {LAW}
 C.T.Law and G.A.Swartzlander, Jr., Chaos Solitons Fractals, 4,
 1759-1766 (1994).
\bibitem {AVC} C.A.Akhmanov, V.A.Vysloukh, A.C.Chirkin, \textit{
 Optics of femtosecond laser pulses}, (Nayka, Moskow, 1988).
\bibitem {FORK}
  Fork R.,Shank C., Hirliman C. et all, Optics
  Lett.,8,1,(1983)
\bibitem {BORN}
 M. Born and E. Wolf, \textit{Principes of Optics}, 6th (corr) ed.,
 (Cambridge University Press, Cambridge, England, 1998).
\bibitem {BERG}
 T. Carozzi, R. Karlsson, and J. Bergman, Phys. Rev. E, 61, 2024-2028
 (2000).
\bibitem {BOYD}
 R.W.Boyd,\textit{ Nonlinear optics}, Acad.Press Inc., (1992).
\bibitem {BAR}
 A.O.Barut, Proc. of the Conf. \textit{ Geometric and Algebraic Aspects
 of Nonlinear Field Theory}(Elsevier,Amsterdam,1989), pp 37-51.
\bibitem {LAN}
 L.D.Landau and E.M.Lifshitz, \textit {Electrodynamics of Continious Media},
 (Nayka, Moskow, 1978).
\bibitem {SHEN}
 Y.R.Shen \textit{The Principles of Nonlinear Optics}, (John Wiley, Sons,
 Inc., 1984).
\end{thebibliography}
\end{document}